\documentclass[aps,prd,preprintnumbers,superscriptaddress,groupedaddress,nofootinbib]{revtex4}

\pdfoutput=1

\usepackage{multirow}
\usepackage{subfigure}
\usepackage{graphicx}
\usepackage{amsfonts}
\usepackage{amsmath,amssymb}
\usepackage{slashed}
\usepackage{feynmp}
\usepackage{caption}
\usepackage{url}
\usepackage{comment}
\usepackage{color}
\usepackage{cancel}
\usepackage{bm}
\usepackage[latin1]{inputenc}
\usepackage{tikz}
\usetikzlibrary{shapes,arrows}

\def\bpm{\begin{pmatrix}}
\def\epm{\end{pmatrix}}

\def \ba {\begin{align}}
\def \ea {\end{align}}

\def \bcom {}

\def \( {\left(}
\def \) {\right)}

\begin{document}

\title{Hiding Thermal Dark Matter with Leptons}

\author{Matthew R.~Buckley}
\affiliation{Department of Physics and Astronomy, Rutgers University, Piscataway, NJ 08854, USA}
\author{David Feld}
\affiliation{Department of Physics and Astronomy, Rutgers University, Piscataway, NJ 08854, USA}

\date{\today}

\begin{abstract}
Any form of dark matter which was in thermal equilibrium with the Standard Model in the early Universe must have some annihilation mechanism in order to avoid overclosure. In general, such models are now constrained by the negative experimental results from colliders, direct detection, and indirect detection, all of which are capable of probing interactions at the approximate strength suggested by a thermal cross section. It is timely to consider what scenarios of thermal dark matter which are still viable. In this paper we consider a class of dark matter models which is designed to avoid many of the current constraints: Majorana dark matter coupling to the Standard Model through leptophilic singlet scalars and pseudoscalars. We show that requiring realistic electroweak symmetry breaking generically forces the mediators to couple with quarks, allowing these models to be constrained by the current experimental data. We find that -- barring fine-tuning -- this type of thermal dark matter is excluded by a combination of direct and indirect detection for masses below $\sim 100$~GeV. Heavier dark matter is still viable, but in principle visible via its indirect detection signature.
\end{abstract}

\maketitle

\section{Introduction}
	
Though the gravitational evidence for dark matter is robust  \cite{Zwicky:1933gu,Rubin:1980zd,Olive:2003iq,Ade:2013zuv}, the particle nature of dark matter has yet to be discovered. Though other possibilities exist, a well motivated paradigm for dark matter is that of a thermal relic, i.e.~a particle which was in thermal equilibrium in the early Universe and whose abundance today is set by the freeze-out of self annihilation. In order for this thermal dark matter annihilation to occur, there must be some lighter particles for dark matter to annihilate into, with a sufficiently large cross section to avoid overclosing the Universe. This type of dark matter is a generalization of Weakly Interacting Massive Particles (WIMPs), where we allow the interaction to be something other than $SU(2)_L\times U(1)_Y$ \cite{Feng:2008ya,Feng:2010gw}. More generally, we can imagine a dark matter model whose abundance is set by some non-thermal mechanism, but which was still in thermal equilibrium in the early Universe. Such dark matter also requires significant annihilation into some lighter state to avoid overclosure; equal to or larger than the thermal cross section \cite{Kolb:1990vq}. 

We can therefore consider ``thermally-coupled dark matter,'' whose overall abundance may or may not be set by freeze-out, but was in equilibrium with the primordial Standard Model bath. The key observation about such thermally coupled dark matter is that -- unlike other possible models of dark matter -- it requires an interaction with some other particle that is known to be approximately the strength of the weak nuclear force or larger. This makes thermally coupled dark matter experimentally accessible, at least in principle.
	
Various experiments are now sensitive to dark matter-Standard Model interactions which are of the approximate strength expected for a thermally-coupled dark matter particle. Different classes of experimental probe orthogonal channels, including scattering off nuclei (direct detection), annihilation in the present-day Universe (indirect detection), and pair production at colliders. Each of these has thus far given a null result (e.g.~\cite{Akerib:2015rjg, Archambault:2017wyh}). With this in mind, an interesting and important question to ask is ``Should we have seen thermal dark matter in these experiments by now?'' Or rather, do models still exist in which dark matter can have sufficiently large interactions with the Standard Model to obtain at least the thermal cross section in the early Universe, but still remain hidden from current direct and indirect detection constraints? If so, what types of models are still viable? 

It is therefore interesting to consider models which are likely to evade the experimental bounds. Many models of thermal dark matter have been put forward \cite{Abdallah:2014hon, Abdallah:2015ter, Abercrombie:2015wmb}, often in the form of simplified models which consider only minimal additional particle content without the complexity of a more complete theory (see the review articles Refs.~\cite{Abdallah:2014hon,Abdallah:2015ter,Abercrombie:2015wmb}). 

In this paper, we are interested in dark matter models which are ``maximally difficult'' to detect, we focus on a model of Majorana dark matter where all interactions with the Standard Model proceed through a gauge-singlet scalar or pseudoscalar mediator \cite{ Buckley:2014fba, Yang:2016wrl,Berlin:2015wwa,Harris:2015kda, Harris:2014hga}. Theoretically, these models are interesting as they can easily be embedded into extended Higgs sectors \cite{LopezHonorez:2012kv,Bell:2016ekl,Bauer:2017ota,Ivanov:2017dad,Casas:2017jjg}. Experimentally, the limits on these models are relatively weak (as compared, for example, to vector-mediator models, see e.g.~\cite{Buchmueller:2014yoa,Fairbairn:2016iuf}), as the coupling to light quarks is expected to be suppressed by the small Standard Model yukawas. Many experimental cross sections then are loop-suppressed and negligible. However, once we require gauge invariance and viable  electroweak symmetry breaking (EWSB), we find new constraints on the particle interactions \cite{ Kahlhoefer:2015bea, Englert:2016joy,Kahlhoefer:2017dnp}, constraints that would not be present when only considering the simplified model itself and which can greatly increasing the experimental sensitivity to the model. 
	
The simplest version of this model is the Standard Model with a minimal Higgs sector \cite{Patt:2006fw,Kim:2006af,MarchRussell:2008yu,Kim:2008pp,Feng:2008mu,Andreas:2008xy,Barger:2008jx,Kadastik:2009ca,Arina:2010an,Kamenik:2012hn,Englert:2011yb,Low:2011kp,Djouadi:2011aa}, plus Majorana dark matter and one additional scalar singlet. Interactions of the dark matter with the Standard Model proceed through mixing of the singlet with the Standard Model Higgs doublet. Though already well considered in the literature, as a precursor to more complicated models, we examine this simple model in detail. To avoid the constraints from direct and indirect searches, along with the requirement that the dark matter be a thermal relic with the correct abundance  ($\Omega_\chi h^2 \approx 0.119$), this simple model is not viable \cite{Escudero:2016gzx}: in order to produce enough dark matter in the early Universe, the effective dark matter-quark interactions (induced by singlet scalar-Higgs mixing) must be large, which in turn would imply dark matter-nuclei interactions which have been ruled out by direct detection. We therefore turn to extensions of this model where the dark matter-quark interactions have been suppressed.
		
Taking the idea of ``quark-phobic'' dark matter to its logical conclusion, we consider a dark matter sector which is ``leptophilic'': preferring to interact with leptons rather than quarks. If this leptophilia can be preserved after EWSB, then direct detection scattering rates will be reduced (by selecting Majorana dark matter and a scalar mediator, we ensure that scattering with quarks occurs only at the the two-loop level, if the mediators coupling only to leptons). Similar difficulties would be expected at hadron colliders, as the colliding protons are comprised primarily of strongly interacting particles. The minimal Standard Model Higgs sector mixed with a new scalar singlet cannot realize leptophilia, but we can write down such a model if we extend the scalar sector. 

We therefore consider a scalar singlet which mixes with a Two-Higgs Doublet Model (2HDM) \cite{Branco:2011iw}, in particular a 2HDM where one of the Higgs doublets couples only to quarks and the other couples only to the leptons \cite{Buckley:2015cia, Grossman:1994jb, Akeroyd:1996he, Aoki:2009ha}. The addition of another Higgs doublet greatly increasing the complexity of both the scalar potential of the model, as well as the Yukawa interactions with fermions. This complexity allows for more freedom in the relative couplings of dark matter to individual fermions. This allows us to impose discrete symmetries allowing one doublet to couple to the leptons, while the other couples to the quarks, at least before EWSB.
	
We study the constraints on this dark matter model from direct/indirect detection, precision Higgs results, and colliders, while requiring a thermal annihilation cross section at least large enough to realize the observed dark matter abundance. Due to the large number of unknown parameters in the most general scalar sector, we scan over the parameter space, requiring a sensible scalar sector (a 125 GeV Standard Model-like Higgs and positive scalar masses squared). We show that, while the 2HDM can be purely leptophilic in the Yukawa sector of the Lagrangian, there exists unavoidable mixing between the quark- and leptophilic Higgses after EWSB. Without fine tuning it is not possible to naturally forbid dark matter-quark interactions while allowing dark matter-lepton interactions. This is an important consequence of the earlier requirement of gauge invariance and viable EWSB. Due to this mixing and the resulting dark matter-quark interaction, we find much of the parameter space for this type of thermally-coupled dark matter is ruled out by existing experimental constraints. 

The indirect detection constraint in particular is very effective at probing the low dark matter mass region, eliminating the majority of points for which $m_\chi \lesssim 100$ GeV. The points that survive the direct detection constraint are tuned to be leptophilic, though approximately $7\%$ of the viable parameter points have a pseudoscalar-top quark couplings  greater than $0.1$, about large as would expected in a ``quarkphilic'' model, and potentially accessible in the full dataset, given the current constraints \cite{Sirunyan:2017hci, ATLAS:2014kua}. In all, we can see that the addition of EWSB constraints to leptophilic scalar mediators makes them far more ``visible'' than naive expectations: the remaining parameter space consists of dark matter which is too heavy to have been seen in indirect detection experiments and interacts via a pseudoscalar mediator with couplings tuned to be leptophilic {\em after} EWSB.

This paper is organized as follows. In Section II we demonstrate our technique by applying the experimental constraints on thermal dark matter to a simple extension of the Standard Model with fermionic dark matter and a scalar singlet, which mixes with the Standard Model Higgs. After showing that this model is ruled out by direct detection, in Section III we consider a leptophilic Two-Higgs Doublet Model, again considering the experimental constraints on this model. We conclude in Section V.

\section{Dark Matter Through the Higgs Portal}

Before studying a leptophilic model, we first consider a simple example, which will demonstrate our approach. We wish to have a Majorana dark matter particle interact via a scalar mediator sector with the Standard Model. Since the Standard Model fermions are chiral and dark matter itself must be electrically neutral, we must have some form of mixing between gauge singlets and $SU(2)_L$ doublets in our model. Either we interact via a $SU(2)_L$ doublet scalar and introduce doublet-singlet mixing in the dark sector, or we have singlet dark matter and a scalar singlet which mixes with a doublet. The former possibility adds additional charged particles to the dark sector which provides additional handles for new physics search, and so we focus on the latter.

We add a complex scalar singlet mediator and Majorana dark matter to the Standard Model, which we assume has the simplest viable Higgs section, consisting of one $SU(2)_L$ doublet. This connect dark matter to the Standard Model through the ``Higgs portal'' \cite{Patt:2006fw,Kim:2006af,MarchRussell:2008yu,Kim:2008pp,Feng:2008mu,Andreas:2008xy,Barger:2008jx,Kadastik:2009ca,Arina:2010an,Kamenik:2012hn,Englert:2011yb,Low:2011kp,Djouadi:2011aa,Escudero:2016gzx}, which is known to be highly constrained from data. To demonstrate the constraints we will consider for the leptophilic model, we reproduce the constraints here.

We require that this model have a self-annihilation rate in the early Universe which is large enough to reduce the thermal relic population to the observed dark matter density or lower (allowing for the possibility of non-thermal dark matter production). We mix our singlet mediator with the Standard Model Higgs doublet; after this mixing we require the scalar spectrum to contain a 125 GeV Higgs boson which is consistent with LHC results.

\subsection{Model}

Adding one complex scalar which is a singlet under all gauge groups to the minimal Higgs sector of the Standard Model results in an expanded scalar potential in which the doublet Higgs $\Phi$ mixes with the singlet $\varphi_S = \phi_S + i a_S$. The most general gauge-invariant potential takes the form: 

\begin{align}
V(\Phi, \varphi_S) = V_2 + V_3 + V_4,
\end{align}
where
\begin{align}
V_2 &= m_1^2 |\Phi|^2  + \frac{1}{2} m_{S_1}^2 | \varphi_S |^2 - \left( \frac{1}{2} m_{S_2}^2 \varphi_S^2 + \mbox{h.c.} \right) \\
V_3 &=    \kappa_1 |\Phi|^2 \varphi_S  +    \kappa_2 |\varphi_S |^2 \varphi_S     +    \kappa_3 \varphi_S^3   +\mbox{h.c.}  \\
V_4 &= \lambda_1 |\Phi|^4 + \lambda_{2} |\Phi|^2 |\varphi_S|^2 +\lambda_{3} |\varphi_S|^4 + \left[ \lambda_4 |\Phi|^2 \varphi_S^2    + \lambda_{5} \varphi_S^4 + \lambda_{6} |\varphi_S|^2 \varphi_S^2  +\mbox{h.c.}\right]  
\end{align}

We will assume the scalar sector is CP-conserving. For real parameters, the mass matrix for the CP-even scalars after electroweak symmetry breaking can be treated as free parameters, trading in the $m_i^2$, $\kappa$, and $\lambda$ parameters in $V$. Note that this model has one CP-odd scalar, which is purely singlet. The CP-odd components of $\Phi$ and $\varphi_S$ do not mix, even after EWSB, since the doublet only enters in the potential through the combination $|\Phi|^2$. One can easily check that $|\Phi|^2$ contains terms proportional to $a_h^2$ (where $a_h$ is the CP-odd component of $\Phi$) but not $a_h$, and thus will never mix with the singlet pseudoscalar. The CP-odd degree of freedom $a_h$ is then eaten by the $Z$ during EWSB, leaving the pseudoscalar component of $\varphi_S$ as the only new CP-odd particle.

In general, the two physical CP-even scalars will be related to the gauge eigenstates through a rotation angle $\theta$:
\begin{align}
h =  \phi \cos \theta  +  \phi_S \sin \theta  \label{eq: h125}\\
H = - \phi \sin \theta + \phi_S \cos \theta ,
\end{align}
where $\phi$ is the CP-even component of the Higgs doublet. We take $h$ to be the 125 GeV Higgs, and $m_H > m_h$. 

The dark matter sector has the Majorana dark matter $\chi$ coupling to the fermions through a Yukawa interaction:
\begin{align} 
{\cal L}_{\rm yuk} \supseteq~& y_\chi \varphi_S \chi \chi +  \frac{{\bf y}_\ell}{\sqrt{2}}  ({\bf L}\cdot \Phi^*)\bar{\bf \ell}_R  + \frac{{\bf y}_u}{\sqrt{2}}  ({\bf Q}\cdot \Phi) \bar{\bf u}_R + \frac{{\bf y}_d}{\sqrt{2}} ({\bf Q}\cdot\Phi^*) \bar{\bf d}_R + \mbox{h.c.} \nonumber
\end{align}
Our normalization convention is chosen such that $\langle \Phi \rangle \approx 246$~GeV and $y_t \sim 1$. 
After electroweak symmetry breaking, the fermions obtain mass and the scalars rotate into their physical states.  As the doublet sector contains only one field, the dark matter-Standard Model interaction will be proportional to fermion mass multiplied by the relevant mixing angle. This makes this model automatically minimally flavor violating \cite{Isidori:2012ts}. The scalar interactions with the gauge bosons similarly get modified by the relevant mixing angle:
\begin{align} 
{\cal L}_{\rm gauge} \supset~& \frac{g c_\theta m_Z}{c_W} h Z Z + g c_\theta m_W h W^+ W^- - \frac{g s_\theta m_Z}{c_W} H Z Z - g s_\theta m_W H W^+ W^- .
\end{align}
Here $s_\theta=\sin\theta$ and $c_\theta= \cos\theta$.

Higgs measurements from the LHC have determined that the 125 GeV Higgs is very nearly Standard Model-like in its couplings. This places a constraint on the mixing parameter: $\cos^2 \theta > 0.88$ at $95 \%$ CL \cite{ATLAS:2014kua}. This gives a range for the combination $|s_\theta c_\theta|$: $ 0 < |s_\theta c_\theta| < 0.32$, which appears in the effective couplings between dark-matter and fermions/gauge bosons. To display constraints, we'll choose the maximal value $|s_\theta c_\theta| = 0.32$ which corresponds to maximal mixing allowed by experiment. The interactions with the dark sector will be proportional to the dark matter Yukawa coupling $y_\chi$ times these mixing angles, so assuming a smaller $|s_\theta c_\theta|$ would translate to a larger $y_\chi$, without changing our relative limits. 

\subsection{Relic Abundance}

The dominant processes contributing to dark matter freezeout in the early Universe are $\chi \chi \to \bar{f} f$ and $\chi \chi \to V V^*$, where $V$ are the electroweak gauge bosons. These processes are mediated by $h /H$, and are $p$-wave suppressed. If annihilation through a pseudoscalar were possible the process would not be velocity-suppressed, however as explained in previously, the CP-odd components of $\Phi$ and $\varphi_S$ do not mix and so there is no pseudoscalar that connects the Standard Model to the dark sector in this model.

Our assumption of thermally-coupled dark matter translates into the requirement that the thermal relic density $\Omega_\chi h^2$ be $0.119$ or smaller, or 
 or equivalently a thermal averaged cross section of $ \langle \sigma v \rangle \geq 2.8 \times 10^{-9}$ GeV$^{-2}$. Explicitly, the thermal cross sections take the form:
\begin{align}
\langle \sigma v \rangle_{\chi \chi \to \bar{f} f} &=   \frac{ 3 N_c m_\chi^2 m_f^2 y_\chi^2 s_\theta^2 c_\theta^2}{16 \pi v^2}   \( 1-\frac{m_f^2}{m_\chi^2} \) ^{3/2} \left( \frac{ 1 }{4 m_\chi^2-m_{h}^2} - \frac{1 }{4 m_\chi^2-m_{H}^2}  \right)^2 (T/m_\chi), \\
\langle\sigma v\rangle_{\chi \chi \to V V^*} &=  \frac{ 3  m_V^4 y_\chi^2 s_\theta^2 c_\theta^2}{2 \pi v^2}   \( 1-\frac{m_V^2}{m_\chi^2} \) ^{1/2} \left( \frac{ 1 }{4 m_\chi^2-m_{h}^2} - \frac{1 }{4 m_\chi^2-m_{H}^2}  \right)^2 (T/m_\chi),
\end{align}
where $N_c$ is the fermion color factor. Note that, as $m_H$ approaches $m_h$, the annihilation cross section decreases, and larger couplings are required to obtain the relic abundance. Thus, in what follows we decouple the heavy Higgs, sending $m_H \gg m_h$. Outside of narrow resonance windows, this choice will not greatly affect the other experimental constraints.

Figure \ref{fig: bounds} shows bounds on the $h \chi \chi$ coupling coming from the requirement that the model obtain a sufficiently large thermal cross section $\langle \sigma v \rangle \geq 2.8 \times 10^{-9}$ GeV$^{-2}$. 
Note the dips in the thermal relic curve are due to new annihilation modes becoming kinematically accessible.

\subsection{Invisible Higgs Width}

The invisible width of the Higgs must be taken into account as well. If the dark matter is sufficiently light, the Higgs can decay into a pair of dark matter particles. The upper limit of the invisible branching ratio is 54\% of the total Higgs width \cite{Chatrchyan:2014tja}. Since the doublet part of the 125 GeV scalar has exactly Standard Model couplings, from Eq.~\eqref{eq: h125} we see that its width will be $\cos^2 \theta\times \Gamma_{SM}$. Also since the invisible width is mediated by the singlet component of $h$, we have $\Gamma_{h \to \chi \chi} = \sin^2 \theta\times  \Gamma_{\phi_S \to \chi \chi}$. Therefore, we have the constraint:

\begin{align}
\sin^2 \theta \times \Gamma_{\phi_S \to \chi \chi}  < 0.54 \times \left( \cos^2 \theta\times \Gamma_{SM} + \sin^2 \theta\times \Gamma_{\phi_S \to \chi \chi}  \right),
\end{align}
where
\begin{align}
\Gamma_{\phi_S \to \chi \chi} = \frac{m_h y_\chi^2 }{8 \pi} \left(1 - \frac{4 m_\chi^2}{m_h^2} \right)^{3/2}
\end{align}
The resulting limit on the dark matter Yukawa coupling is shown in Figure~\ref{fig: bounds}.
Note that this measurement constrains $y_\chi \tan \theta$, rather than the combination $y_\chi \sin\theta\cos\theta$, which all other experimental searches are sensitive to. Thus, changing our assumption of $|s_\theta c_\theta| = 0.32$ could strengthen (weaken) this limit relative to the others as $\tan \theta$ increases (decreases). As this constraint subdominant, this ambiguity does not materially affect the conclusion.

\subsection{Direct Detection}
Direct detection experiments aim to measure recoil energy from dark matter scattering of nuclei. This model allows for spin-independent direct detection from dark matter scattering off nuclei through an exchange of $h/H$. The dark matter-nucleon cross section is given by \cite{Buckley:2013jwa}:
\begin{align}
&\sigma_{\chi-n} = \frac{4 \mu^2} {\pi} f^2, \\
&f = \sum_{q = u,d,s} \xi_q f_q \frac{m_p}{m_q}  + \frac{2}{27} f_{TG} \sum_{q = c,b,t} \xi_q \frac{m_p}{m_q},
\end{align}
where $\xi_q$ is the effective dark matter quark coupling, $f_{q/TG}$ are nuclear form factors,  and $m_p$ is the proton mass. The effective couplings for our model are:
\begin{align}
\xi_{q} = \frac{m_q y_\chi s_\theta c_\theta}{v } \left( \frac{1 }{m_{h}^2} - \frac{ 1 }{m_{H}^2}   \right).
\end{align}

\begin{figure}[h!]
\includegraphics[width=0.75\columnwidth]{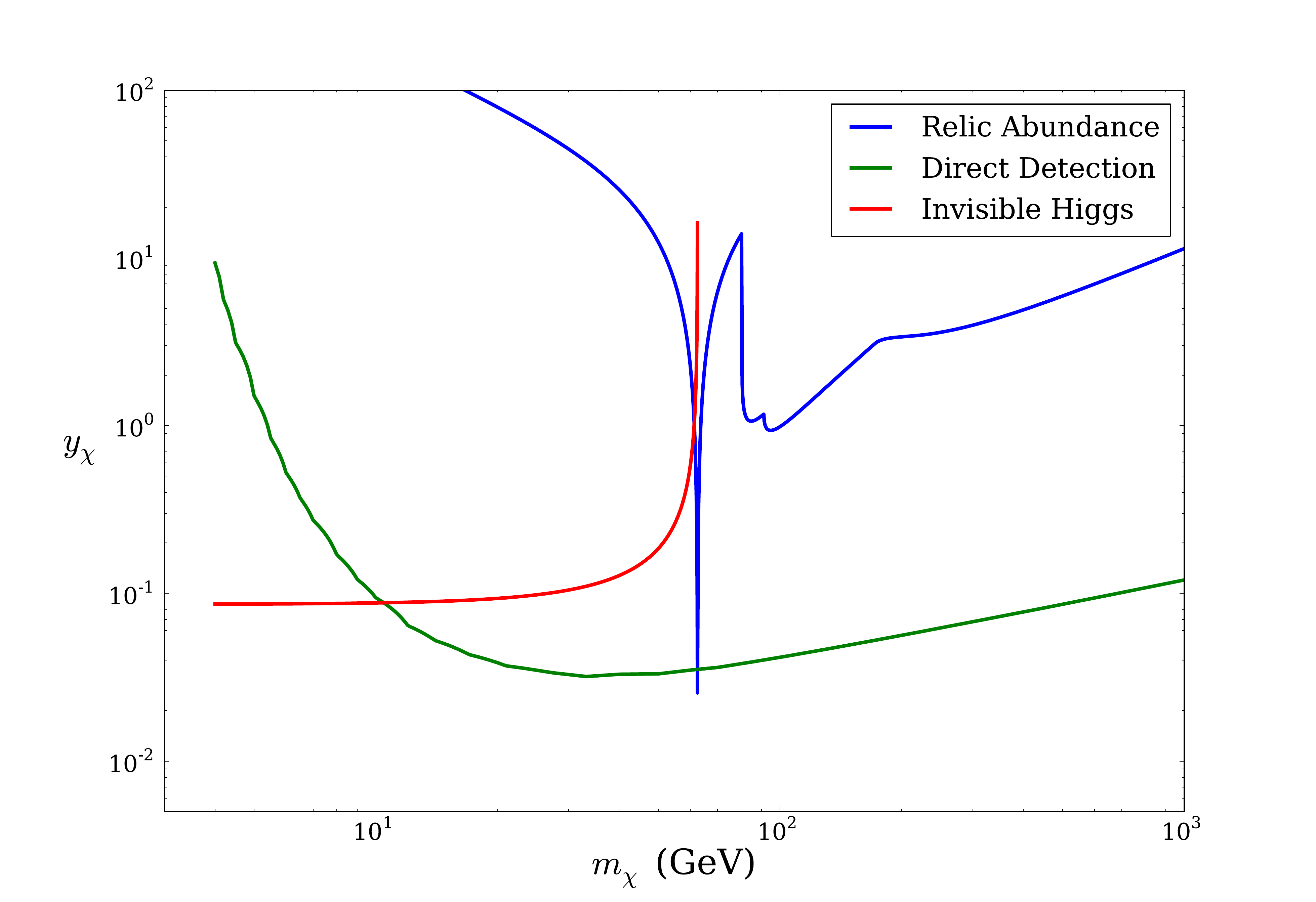}
\caption{Allowed parameter space for the $h \chi \chi$ coupling $ y_\chi$, under the assumption $|s_\theta c_\theta| =0.32$. The blue line indicates the minimum coupling allowed to obtain at least the thermal relic cross section, while the upper limits on this coupling are set by LUX \cite{Akerib:2015rjg} (green), and measurements of the invisible Higgs branching ratio \cite{CMS-PAS-HIG-14-002,Khachatryan:2014iha} (red). Other than a narrow window for $m_\chi \approx (125~\mbox{GeV})/2$, no viable parameter space exists. }
\label{fig: bounds}
\end{figure}

In the mass range of interest, the most stringent bound on the spin-independent cross section comes from the LUX experiment \cite{Akerib:2015rjg}. Bounds coming from direct detection are shown in Figure \ref{fig: bounds} in the limit $m_H \gg m_h$. Clearly, satisfying direct detection constraints is incompatible with obtaining the correct relic abundance. Bringing $m_H$ down towards $m_h$ can relax these limits by ${\cal O}(1)$ factors at best, barring accidental cancellations in the relic abundance. 

Therefore, barring extreme fine-tuning, we can see that the single-Higgs version of the scalar mediator model is completely excluded by the experimental searches -- primarily the lack of signal in direct detection. In order to have a large enough coupling with the Standard Model to annihilate efficiently in the early Universe, this simple model would also have couplings with quarks which are far too large to evade detection in LUX.

\section{A Leptophilic Model}

Given the failure of the simplest model of a scalar mediator, we are led to consider the next simplest case: a Two-Higgs doublet model (2HDM) plus a scalar singlet. The requirement of gauge invariance will dictate the form of the scalar potential and constrains parameters to successfully reproduce Standard Model electroweak symmetry breaking. As the sticking point of the simplest Higgs model was the large direct detection cross section, set by the mediator coupling to quarks, we are motivated to consider the ``leptophilic'' 2HDM in which one scalar doublet couples to quarks and the other to leptons (see e.g.~\cite{Branco:2011iw} for a review of Two-Higgs Doublet Models). As the sensitivity of dark matter direct detection to lepton couplings is highly suppressed, we would expect a 2HDM where both Higgses had couplings to quarks to be more easily detected. Thus, our leptophilic model is the scalar mediator model most likely to evade experimental constraints while still realizing viable thermally coupled dark matter.

\subsection{Model}
The scalar sector consists of two $SU(2)_L$ doublets and a singlet. A general 2HDM leads to tree-level flavor changing neutral currents (FCNC). To avoid this problem, it is common to impose discrete symmetries on the doublets such that they couple only to certain flavors of fermions \cite{Branco:2011iw}. 
As discussed, we work in the leptophilic 2HDM for the remainder of this work. Before EWSB, $\Phi_1$ is the the quark-coupled Higgs and $\Phi_2$ is the lepton-coupled Higgs, which mix with the complex scalar singlet $\varphi_S$. 

In general all scalars can acquire vacuum expectation values (vevs), and the two doublets can be written as:
\begin{align}
\Phi_i =  
 \begin{pmatrix}
  \phi^+_i   \\
  1/\sqrt{2}  \( \phi^0_i + i A_i + v_i  \)
 \end{pmatrix} _{Y = + 1/2}
\end{align}
The Standard Model EWSB vev is given by $v^2 = v_1^2 +v_2^2$. These fields result in a complicated scalar potential:
\begin{align}
V(\Phi_1, \Phi_2, \varphi_S) = V_2 + V_3 + V_4,
\end{align}
where
\begin{align}
V_2 &= m_1^2 |\Phi_1|^2 +m_2^2 |\Phi_2|^2 +1/2 m_{S_1}^2 | \varphi_S |^2 - \left(  \mu_{12}^2  \Phi_1^\dagger \Phi_2 + \frac{1}{2} m_{S_2} \varphi_S^2 + \mbox{h.c.} \right) \\
V_3 &=    \kappa_1 |\Phi_1|^2 \varphi_S    +   \kappa_2 |\Phi_2|^2 \varphi_S     +   \kappa_3   \Phi_1^\dagger \Phi_2 \varphi_S     +   \kappa_4   \Phi_1^\dagger \Phi_2 \varphi_S^\dagger   +    \kappa_5 |\varphi_S |^2 \varphi_S     +    \kappa_6 \varphi_S^3   + h.c.  \\
\begin{split}
V_4 &= \lambda_1 |\Phi_1|^4 + \lambda_2 |\Phi_2|^4 + \lambda_3 |\Phi_1|^2 |\Phi_2|^2 + \lambda_4 |\Phi_1^\dagger \Phi_2|^2 + \Big[ \frac{\lambda_5}{2}    (\Phi_1^\dagger \Phi_2)^2   + \lambda_6   |\Phi_1|^2 (\Phi_1^\dagger \Phi_2)     + \lambda_7   |\Phi_2|^2 (\Phi_1^\dagger \Phi_2)    + \lambda_8 |\Phi_1|^2 \varphi_S^2  \\
&+ \lambda_{9} |\Phi_2|^2 \varphi_S^2 +  \lambda_{10}   (\Phi_1^\dagger \Phi_2) \varphi_S^2  +  \lambda_{11}   (\Phi_1^\dagger \Phi_2) \varphi_S^{\dagger 2}     + \lambda_{12} \varphi_S^4 + \lambda_{13} |\varphi_S|^2 \varphi_S^2  + \lambda_{17} (\Phi_1^\dagger \Phi_2)  |\varphi_S|^2  + \mbox{h.c.} \Big] \\
 & + \lambda_{14} |\Phi_1|^2 |\varphi_S|^2 + \lambda_{15} |\Phi_2|^2 |\varphi_S|^2 +\lambda_{16} |\varphi_S|^4 \nonumber
\end{split}
\end{align}

The assumption of CP conservation requires all parameters in the potential to be real. The requirement of successful EWSB places constraints on parameters by solving for the EWSB minimization and stability conditions. These conditions can easily be used to eliminate $m_1$, $m_2$, and $m_S$. The mass matrices are more conveniently expressed if we define five dimensionful parameters $\rho_{i}$. These parameters and other important notation are described further in Appendix A.  The resulting physical states are comprised of three CP-even scalars $\phi_i$ and two CP-odd scalars $a_i$, where we take $\phi_1 \equiv h$ to be the 125 GeV Higgs. 

The Yukawa Lagrangian takes the form:
\begin{align}
\begin{split}
{\cal L}_{\rm Yuk} \supseteq~  & \sum_{f= u, d, \ell} \frac{m_f }{ v \xi_f^\phi }  N_{fi}  \phi_i  \bar{f} f  - i \frac{m_f }{ v \xi_f^A}  A_{fi}  a_i  \bar{f} \gamma^5 f  + y_\chi N_{3i} \phi_i  {\chi}  \chi + i y_\chi A_{3i} a_i {\chi} \gamma^5 \chi
\end{split}
\end{align}
where $i$ is implicitly summed over $i = 1 ,2, 3$ and $\xi_q^\phi \equiv s_\beta$, $\xi_\ell^\phi \equiv c_\beta$, $\xi_q^A \equiv t_\beta$, and $\xi_\ell^A \equiv t_\beta^{-1}$ . The coefficients $N_{ij}$ and $A_{ij}$ come from the unitary matrices that diagonalize the mass matrices (see Appendix A). Here $N_{fi}=N_{1i} (N_{2i})$ for quarks (leptons). The gauge interactions become:
\begin{align} 
{\cal L}_{\rm gauge} \supseteq~& \frac{m_Z^2}{2 v \( s_\beta + c_\beta \) }  \( N_{1i} + N_{2i} \) \phi_i Z Z +  \frac{m_W^2}{v \( s_\beta + c_\beta \) } \( N_{1i} + N_{2i} \) \phi_i W^+ W^- .
\end{align}

LHC Higgs measurements have determined that the 125 GeV Higgs has Standard Model like couplings \cite{TheATLASandCMSCollaborations:2015bln}. This translates to the requirement that $N_{11} + N_{21} \approx s_\beta + c_\beta$. In a 2HDM without a singlet, $N_{11} = c_\alpha, N_{21} = -s_\alpha$ where $\alpha$ rotates the CP-even states into physical states. Thus, the requirement that the 125 GeV Higgs be nearly Standard Model-like in its coupling to gauge bosons reproduces the standard alignment limit result of $\alpha = \beta - \pi/2$ \cite{Carena:2013ooa, Carena:2015moc}.

To ensure stability of the dark matter, we charge it under some global symmetry. To allow a $\varphi_S \chi \chi$ term, we can charge $\chi$ under a ${\mathbb Z}_2$, or we can charge both $\varphi_S$ and $\chi$ under a ${\mathbb Z}_3$. If we assume that $\varphi_S$ is charged under some ${\mathbb Z}_3$ (such that the term $\varphi_S \chi \chi$ is neutral) then the following terms on the potential must be zero: $ \kappa_1, \kappa_2,  \kappa_3, \kappa_4, \kappa_5, \lambda_8, \lambda_9,  \lambda_{10},  \lambda_{11},  \lambda_{12}, \lambda_{13},$ and $m_{S_2}$. 

An interesting consequence of this symmetry is that the dark matter would not communicate with the Standard Model through the pseudoscalar, as the parameter $\rho_5$ (built from the couplings and masses in the unbroken phase) would be zero. As seen in Appendix A, this controls the mixing in the singlet-doublet pseudoscalar mass matrix. As in the single Higgs model considered previously, without this mixing the thermal freeze-out cross sections would all be $p$-wave suppressed, requiring large Yukawas to overcome. Again following our motivation to study the thermally-coupled dark matter model which is ``maximally difficult'' to discover, we focus on the case where $\varphi_S$ is uncharged under a ${\mathbb Z}_3$ symmetry. This allows for $s$-wave annihilation of dark matter into Standard Model fermion, and thus smaller couplings which can more easily evade the experimental searches.

However, this choice means that, while we can make our scalars leptophilic while in gauge eigenstates, we cannot impose a symmetry that bars mixing between the scalar singlet and the CP-even components of {\em both} the doublets. This is because there is no way to forbid the terms $| \Phi_1 |^2 |\varphi_S |^2$ and $| \Phi_2 |^2 |\varphi_S |^2$ in the potential. If $\varphi_S$ is neutral under all symmetries then it would not necessarily need to acquire a vev for a viable mass spectrum (if $\varphi_S$ is charged it must acquire a vev to give mass to $\chi$) and so these two terms would not cause mixing, but the $ | \Phi_1 |^2 \varphi_S$ and $ | \Phi_2 |^2 \varphi_S$ terms would. Furthermore,  the leptophilic limit is not achieved by large $\tan \beta$, as in this limit the $\Phi_1 - \phi_S$ mixing term goes like the parameter $ \rho_3$. Therefore, to achieve an approximately leptophilic model, we would additionally have to require that $\rho_3$ be small, not just large $\tan\beta$. In all cases, leptophilic couplings after EWSB occur only when couplings which are allowed by symmetries are unnaturally small.

We are now interesting in discovering the experimental limits on the parameters of this model. However, after the minimization conditions and the redefinition of parameters (Appendix A), we have reduced the effective parameter space from 29 free parameters to 14: ($\rho_1 - \rho_5, \lambda_1, \lambda_2, \lambda_{345}, \lambda_5, \lambda_6, \lambda_7, m_{33}, m_{a_{33}}, \tan \beta$). A common assumption in 2HDM is that discrete symmetries in the Higgs sector prohibit the quartic terms odd in either doublet (though this symmetry is softly broken by quadratic terms \cite{Branco:2011iw}). Since this simplifies the form of the mass matrices without greatly affecting the phenomenology, we will make this choice as well and set $\lambda_6 = \lambda_7 = 0$. This is still a large number of parameters, and to cover it, we scan over the multidimensional parameter space. 

\begin{figure}[h!]
\includegraphics[width=1.1\columnwidth]{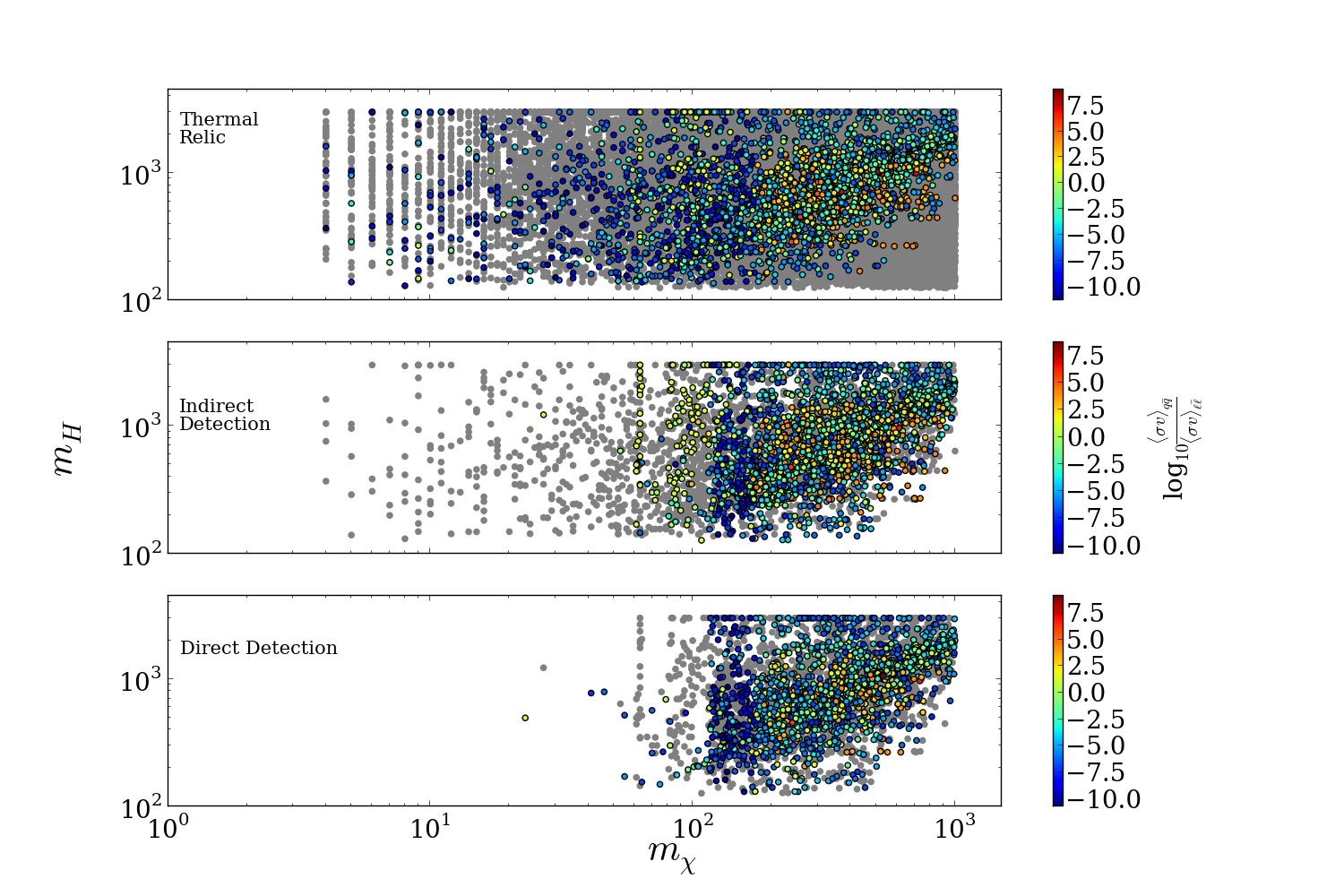}
\caption{\textbf{Top:} All points scanned (grey) and points that give the canonical thermal annihilation cross section (color) in the $m_\chi-m_H$ plane. \textbf{Center:} All points giving canonical $\langle \sigma v \rangle$ (grey) and points that also pass indirect detection constraints (color) in the $m_\chi-m_H$ plane. \textbf{Bottom:} All points passing the thermal and direct detection constraints (grey) and points that also survive direct detection constraints (color) in the $m_\chi-m_H$ plane. Colorbar represents the ratio of dark matter thermal annihilation to quarks and leptons.}
\label{fig: mchimH}
\end{figure}

\begin{figure}[h!]
\includegraphics[width=1.1\columnwidth]{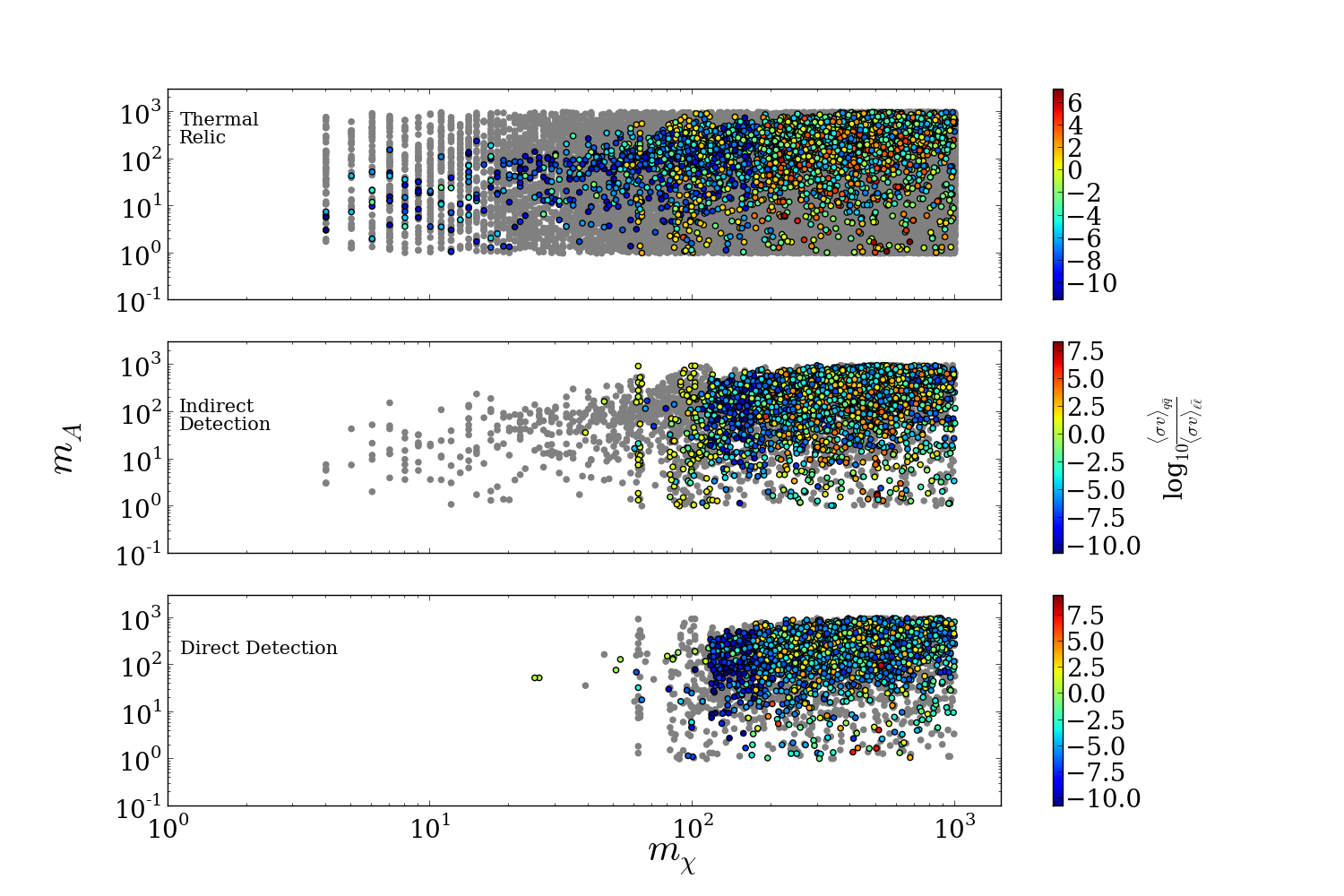}
\caption{\textbf{Top:} All points scanned (grey) and points that give the canonical thermal annihilation cross section (color) in the $m_\chi-m_A$ plane. \textbf{Center:} All points giving canonical $\langle \sigma v \rangle$ (grey) and points that also pass indirect detection constraints (color) in the $m_\chi-m_A$ plane. \textbf{Bottom:} All points passing the thermal and direct detection constraints (grey) and points that also survive direct detection constraints (color) in the $m_\chi-m_A$ plane. Colorbar represents the ratio of dark matter thermal annihilation to quarks and leptons.}
\label{fig: mchimA}
\end{figure}

\begin{figure}[h!]
\includegraphics[width=1.1\columnwidth]{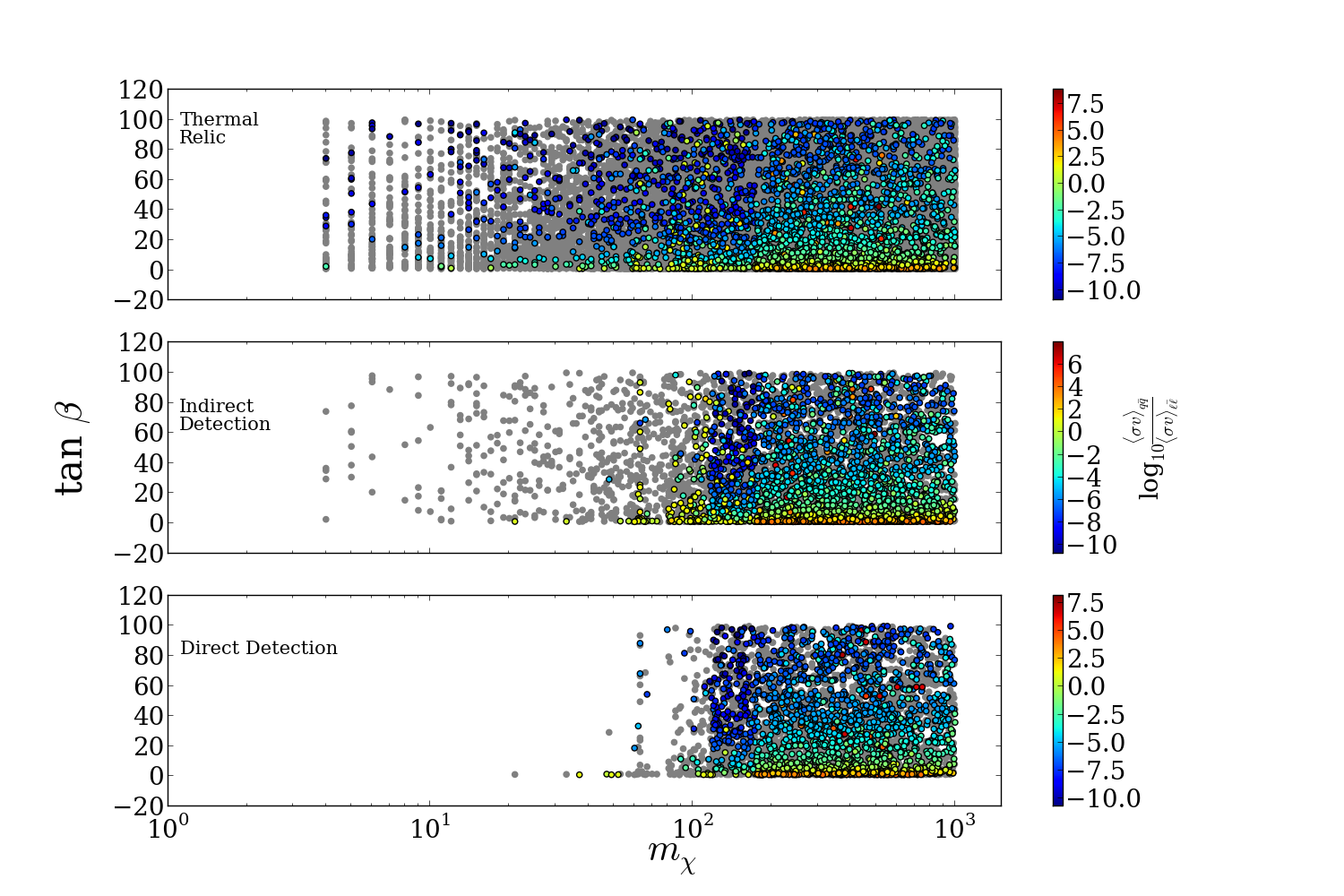}
\caption{\textbf{Top:} All points scanned (grey) and points that give the canonical thermal annihilation cross section (color) in the $m_\chi-\tan \beta$ plane. \textbf{Center:} All points giving canonical $\langle \sigma v \rangle$ (grey) and points that also pass indirect detection constraints (color) in the $m_\chi-\tan \beta$ plane. \textbf{Bottom:} All points passing the thermal and direct detection constraints  (grey) and points that also survive direct detection constraints (color) in the $m_\chi-\tan \beta$ plane. Colorbar represents the ratio of dark matter thermal annihilation to quarks and leptons.}
\label{fig: mchitanb}
\end{figure}

A scan of $10^6$ points was performed over the  parameters $(\rho_1 - \rho_5, \lambda_1, \lambda_2, \lambda_{345}, \lambda_5, m_{a_{33}}^2, \tan \beta$). At each point, the dark matter coupling $y_\chi$ is set by the requirement that the model contain thermal dark matter, as we will describe. The $\rho$ parameters along with $m_{a_{33}}^2$ were scanned in the range: [1 GeV$^2$,1 TeV$^2$], while the $\lambda$ parameters ranged from $[10^{-2}, 10]$, and $\tan \beta$ ranged from [1,100]. For simplicity, the mass of the heaviest scalar $m_{33}$ was set to 3 TeV, as this scalar does not greatly affect the dark matter phenomenology. Initial screening of points require the correct Higgs mass, positive eigenvalues of scalars, and the alignment condition to be met. We then apply the experimental constraints, as described below. The results of the progressive application of the experimental constraints are summarized in Figures \ref{fig: mchimH} -- \ref{fig: mchitanb}, which display the surviving scan points plotted against the physical masses and mixing angles. 

\subsection{Relic Abundance}

This model has both scalars and pseudoscalars connecting Standard Model particles to the dark sector. Since pseudoscalar-mediated annihilation is not $p$-wave suppressed, these processes will be the dominant contribution to dark matter annihilation over the scalar-mediated contribution. As explained previously, this was a motivation for choosing $\varphi_S$ to be uncharged under a discrete symmetry. The cross sections take the form: 
\begin{align} \label{eq: TR1}
\langle \sigma v \rangle_{\chi \chi \underrightarrow{\phi} ff} &=  \sum_f \frac{ 3 N_c m_\chi^2 m_f^2 y_\chi^2 }{16 \pi  \left( v \xi^\phi_{q / \ell} \right)^2}   \( 1-\frac{m_f^2}{m_\chi^2} \) ^{3/2} \left( \frac{N_{i1} N_{31} }{4 m_\chi^2-m_{\phi_1}^2} + \frac{N_{i2} N_{32} }{4 m_\chi^2-m_{\phi_2}^2} +  \frac{ N_{i3} N_{33} }{4 m_\chi^2-m_{\phi_3}^2}  \right)^2 (T/m_\chi) \\
\label{eq: TR2}
\langle\sigma v\rangle_{\chi \chi \to V V^*} &=  \frac{ 3  m_V^4 y_\chi^2}{2 \pi v^2 \left( s_\beta + c_\beta \right)^2}   \( 1-\frac{m_V^2}{m_\chi^2} \) ^{1/2} \left[ \sum_{k = 1,2,3}  N_{3k} \left( \frac{ N_{1k} + N_{2k} }{4 m_\chi^2-m_{\phi_k}^2} \right) \right]^2 (T/m_\chi) \\
\label{eq: TR3}
\langle\sigma v\rangle_{\chi \chi \underrightarrow{A} ff} & =  \sum_{f}  \frac{ N_c  m_f^2}{2 \pi \left( v \xi^A_{q / \ell} \right)^2 }  \left(  \frac{A_{i2} A_{32} }{4 m_\chi^2-m_{a_1}^2} +  \frac{ A_{i3} A_{33} }{4 m_\chi^2-m_{a_2}^2}  \right)^2 \left [m_\chi^2 \sqrt{1-\frac{m_f^2}{m_\chi^2}} +\frac{3m_f^2}{4 m_\chi \sqrt{1-\frac{m_f^2}{m_\chi^2}} } T \right]
\end{align}
where $N_c$ is a fermion color factor, $\xi_q^\phi \equiv s_\beta$, $\xi_\ell^\phi \equiv c_\beta$, $\xi_q^A \equiv t_\beta$, and  $\xi_\ell^A \equiv t_\beta^{-1}$, and $i = 1,2$ for quarks and leptons respectively.

For each parameter point in the scan, we select $y_\chi$ to be the minimum value which will realize a thermal annihilation cross section. Larger values of $y_\chi$ would of course reduce the dark matter relic abundance below the thermal value, which could be desired for models with non-thermal production mechanisms. However, this would also strengthen all other experimental constraints on the model. We eliminate a scan point if the required minimum value of $y_\chi$ is nonperturbative, $y_\chi^2 \geq 4\pi$.

Figures \ref{fig: mchimH} -- \ref{fig: mchitanb} show scatter plots of the scan points where constraints are progressively introduced to display their affect on the parameter space. Figure~\ref{fig: mchimH} is a scatter plot of dark matter mass versus the scalar mediator mass, Figure~\ref{fig: mchimA} is dark matter mass versus pseudoscalar mediator mass, and Figure~\ref{fig: mchitanb} is dark matter mass versus $\tan\beta$.

The effect of the constraint requiring the correct relic abundance is shown in the top panels of each figure. A significant portion of the scanned parameter space is eliminated from this constraint alone, with only 6\% of the points surviving. This validates our choice to select the $\mathbb{Z}_2$ symmetry which allows pseudoscalar mediation, as a scalar mediator would generically require even larger couplings. The figures also feature color shading corresponding to the contribution of quark/lepton annihilation to $\langle \sigma v \rangle$ in order to quantify the ``leptophilic-ness" of the model. Indeed, we see from these figures that the majority of acceptable points are ``leptophilic'' by this measure, though a significant cross section into quarks is present in many points, especially if $m_\chi$ is larger than the mass of the top. This is due to the large top Yukawa, which allows annihilation into tops to remain competitive even with a suppression of the $\varphi_S-\Phi_1$ mixing angle.

\subsection{Indirect Detection}
Indirect detection experiments look for the results of dark matter annihilation in the Universe today. Such processes could be detected by observing an excess of gamma rays or positrons coming from an area of high dark matter density. The temperature of the Universe today is negligible and so only annihilation cross sections which are temperature independent can contribute to the indirect detection of dark matter. From Eqs.~(\ref{eq: TR1} --\ref{eq: TR3}), we see that only the pseudoscalar can mediate such processes. The best current constraints come from the stacked dwarf galaxy search by the {\em Fermi} Large Area Telescope ({\em Fermi}-LAT) \cite{Archambault:2017wyh}, which is sensitive to the gamma rays that would be emitted by the decay of the heavy fermions produced in these annihilations. Specifically the $\chi \chi \to \tau \tau$ process sets the tightest limits for our model.

The importance of these constraints can be seen in the bottom panels of Figures \ref{fig: mchimH}--\ref{fig: mchitanb}. Nearly all points with $m_\chi \lesssim 100$ GeV are removed, with the exception of resonance points. As more data is collected from the Fermi-LAT and more targets of high dark matter density are identified \cite{Fermi-LAT:2016uux}, it is possible that indirect detection could rule out even more of this parameter space.

\subsection{Direct Detection}
Direct detection constraints will come from spin-independent interactions mediated by the CP even scalars. The effective dark matter-quark coupling in this model is:

\begin{align}
\xi_{q} = \frac{m_q y_\chi}{v s_\beta} \left( \frac{N_{11} N_{31} }{m_{\phi_1}^2} + \frac{N_{12} N_{32} }{m_{\phi_2}^2} +  \frac{ N_{13} N_{33} }{m_{\phi_3}^2}  \right)
\end{align}

The reduced mass squared $\mu^2$ takes values between $0.73 - 0.88$~GeV$^2$ for $10 < m_\chi < 1000$ GeV. In this mass range, the upper limits on the cross section from LUX are between $\sim 10^{-18} - 10^{-17}$ GeV$^{-2}$ \cite{Akerib:2015rjg}. This leads to the general requirement that:
\begin{align} \label{eq:22}
\frac{y_\chi}{s_\beta}  
 \left( \frac{N_{11} N_{31} }{m_{\phi_1}^2} + \frac{N_{12} N_{32} }{m_{\phi_2}^2} +  \frac{ N_{13} N_{33} }{m_{\phi_3}^2}  \right) \lesssim 10^{-6} ~\mbox{GeV}^{-2}.
 \end{align}

The effect of the direct detection constraint is shown in the bottom panel of Figures \ref{fig: mchimH}--\ref{fig: mchitanb}. Approximately 3.6\% of points scanned were able to give a thermal relic while avoiding direct/indirect detection constraints. Thus, even though these models were designed to be ``leptophilic'' (and thus one would expect little or no scattering between dark matter and nuclei), after EWSB, we find in general significant mixing and thus tight limits from direct detection. Only those models with relatively heavy dark matter (and thus weaker direct detection bounds) or accidental tunings (resulting in a leptophilic scalar mediator after EWSB) survive.

\begin{figure}[h!]
\includegraphics[width=1.1\columnwidth]{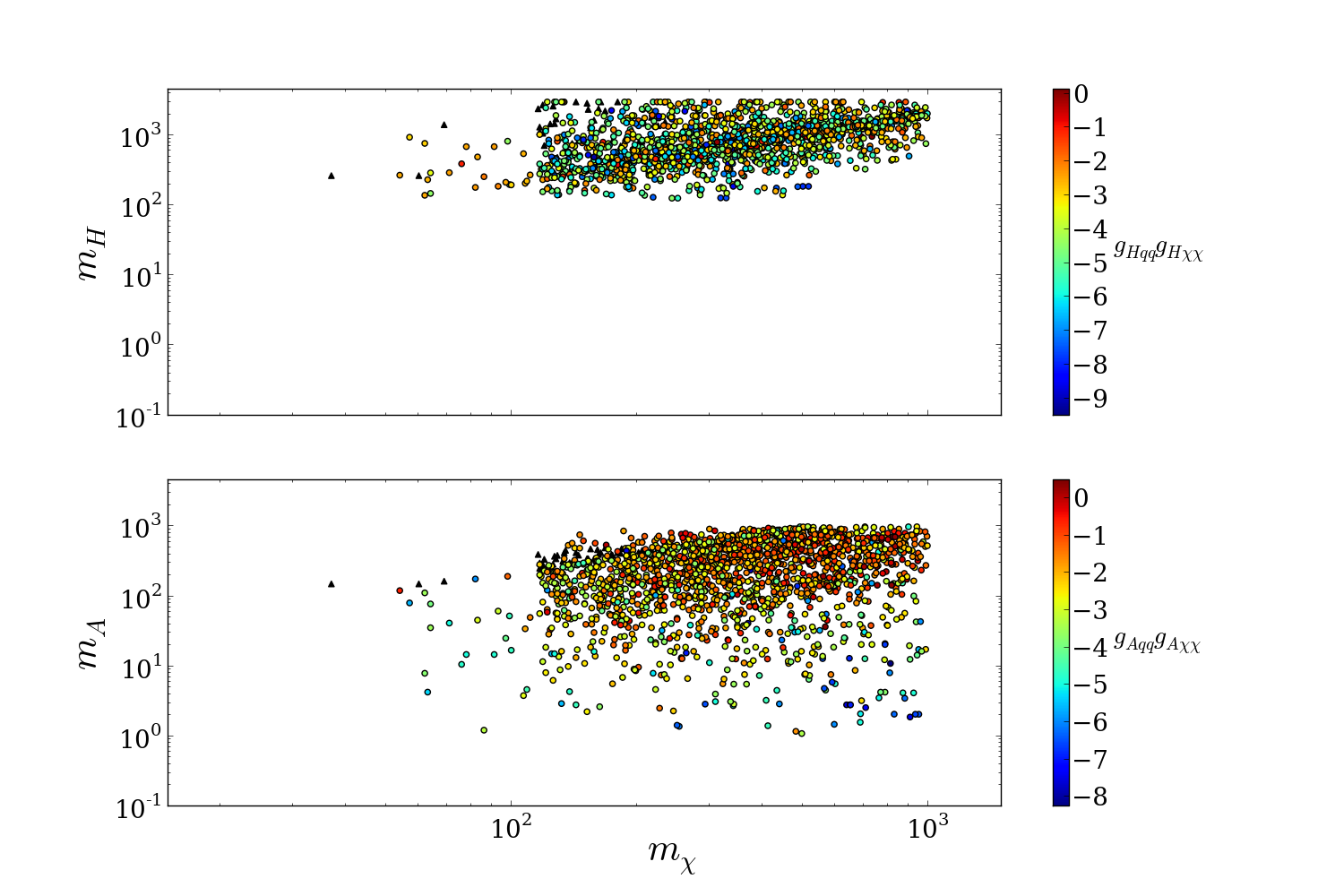}
\caption{Effective quark-dark matter coupling through scalar (top panel) and pseudoscalar (bottom panel). Black triangles are points that are ruled out by CMS \cite{Sirunyan:2017hci}.}
\label{fig: gSMgDM}
\end{figure}

Figure \ref{fig: gSMgDM} shows values of the effective quark-dark matter coupling through both the pseudoscalar and scalar mediator 
\begin{align}
g_{\rm eff} &\equiv g_{Hqq} g_{H \chi \chi} = (y_q N_{12} / s_\beta) (y_\chi N_{32}) \quad \quad \quad (\text{scalar}) \\ 
g_{\rm eff} &\equiv g_{aqq} g_{a \chi \chi} = (y_q A_{12} / t_\beta) (y_\chi A_{32}) \quad \quad \quad (\text{pseudoscalar}).
\end{align}
These effective couplings correspond to the couplings that are relevant to production of dark matter through the mediator at the LHC through top-quark loops.

The present bounds set by CMS and ATLAS~\cite{Sirunyan:2017hci, ATLAS:2014kua} are sensitive to couplings $\gtrsim 1$ for mediators  with masses below a few hundred GeV and the dark matter less than half the mass of the mediator. Very few of the surviving parameter points are eliminated by the current constraints, as shown in Figure \ref{fig: gSMgDM}. However, 7\% of the parameter points that have survived all previous experimental constraints have couplings between the pseudoscalar mediator and top quarks which are greater than $0.1$, equivalent to the strength one would expect for thermal dark matter mediated by a scalar coupling directly to top quarks \cite{Buckley:2014fba}. This subset of ``leptophilic'' models is therefore not particularly less visible than the equivalent models which have couplings to quarks before EWSB.  The inability of the LHC to yet exclude those leptophilic points is not due to the lack of couplings between the mediators and the quarks before EWSB, but simply that it is difficult for the LHC to experimentally exclude scalar or pseudoscalar-mediated dark matter despite ${\cal O}(0.1-1)$ couplings to the top. Coupling only to leptons provides very little advantage in hiding thermal dark matter in this case.

\section{conclusion}

We have studied a class of models in which scalar/pseudoscalar mediators couple gauge singlet Majorana dark matter to the Standard Model. Furthermore we demanded that these models be consistent with gauge invariance above the scale of electroweak symmetry breaking. Scalar mediators are well motivated since many extensions of the Standard Model involve an extended Higgs sector. The simplest case to study is that of the SM with an additional gauge singlet. This model is shown to be unable to conform to current direct detection constraints while simultaneously yielding a thermal relic with the correct abundance. The difficulty arises from the inability to suppress quark-dark matter interactions for the sake of direct detection while keeping the thermal annihilation cross section large enough.

The next model to consider is then a 2HDM with an additional scalar singlet. Due to the difficulties of the previous model, we choose to study the leptophilic 2HDM in which one doublet couples to leptons while the other couples to quarks. One might expect that the quark-dark matter couplings can be small enough to avoid direct detection, while dark matter annihilation through the pseudoscalar to leptons can yield a thermal relic. The pseudoscalar mediated annihilation channel is not velocity suppressed, as opposed to the case with the CP-even scalars. As such, this model is the quintessential example of thermal dark matter which we would expect is nearly undetectable in dark matter experiments.

Constraints from indirect detection and colliders are considered in addition to direct detection and thermal abundance. Our method for analyzing this model was to perform a scan over the multidimensional parameter space. The parameter space is particularly large due to the fact that we consider all terms consistent with gauge invariance. Many of these terms come from the extended scalar potential. After scanning over the parameter space, we find only $\sim 3.6\%$ of points survive both the requirement for a thermal relic abundance of dark matter and all other experimental constraints. The combination of requiring a thermal annihilation cross section and the direct detection limits forces the surviving parameter points to be mostly characterized by a leptophilic pseudoscalar. This requires unusually small couplings which are not set to zero by any symmetry of the model. The low mass dark matter range ($m_\chi \lesssim 100$ GeV) is eliminated by indirect detection constraints from {\em Fermi}-LAT. 

A small percentage of the parameter points with dark matter masses above the {\em Fermi}-LAT limit have couplings to top quarks which are large enough to be potentially accessible at the high-luminosity LHC. For the remaining points, discovery or exclusion will require future improvements in indirect detection limits (perhaps through discovery of new dark matter-rich targets). We can therefore see that, while thermal dark matter mediated by ``leptophilic'' spin-0 mediators are not completely eliminated, the requirements of successful electroweak symmetry breaking greatly constrain the possible parameter space. An ``average'' set of parameters in this model would either fail to obtain a thermal cross section or fail to be sufficiently leptophilic to avoid direct detection; only versions of the model which have a nearly leptophilic pseudoscalar {\em after} EWSB can avoid all the constraints, and this tuning arising accidentally, rather than from some fundamental symmetry of the model.

\appendix
\section{\\Additional Formulae for Section III}

Here we give explicit formulae for the mass matrices of the scalars and pseuodoscalars described in section III. 
The neutral CP even scalars mix in a $3 \times 3$ mass matrix whose components in the $(\phi_1^0, \phi_2^0, \varphi_S)$ basis are:

\begin{align}
m_{11}^2  &=  \rho_1/ t_\beta + 2 v^2 s_\beta^2 \lambda_1 + \frac{3}{2} v^2 s_\beta c_\beta \lambda_6 - \frac{1}{2} v^2 c_\beta^2  \lambda_7 / t_\beta \\
m_{12}^2 &=   -\rho_1  +v^2 s_\beta c_\beta \lambda_{345}  + \frac{3}{2} v^2 s_\beta^2 \lambda_6  + \frac{3}{2} v^2 c_\beta^2 \lambda_7 \label{eq: m12} \\
m_{13}^2 &=  c_\beta \rho_2 + s_\beta \rho_3   \label{eq: m13} \\ 
m_{22}^2 &=  t_\beta \rho_1 + 2 v^2 c_\beta^2 \lambda_2 - \frac{1}{2} v^2 s_\beta^2 t_\beta \lambda_6 + \frac{3}{2} v^2 s_\beta  c_\beta \lambda_7  \\ 
m_{23}^2 &= s_\beta \rho_2 + c_\beta \rho_4  \label{eq: m23} \\
m_{33}^2 &= m_{33}^2,
\end{align}
and the CP odd mass matrix components in the $(a_1, a_2, a_S)$ basis are:

\begin{align}
m_{a _{11}}^2  &= \rho_1 / t_\beta - v^2 c_\beta^2 \lambda_5 - \frac{1}{2} v^2 s_\beta c_\beta \lambda_6  - \frac{1}{2} v^2 c_\beta^2 \lambda_7 / t_\beta   \\
m_{a _{12}}^2 &=  -\rho_1  +v^2 s_\beta c_\beta \lambda_{5}  + \frac{1}{2} v^2 s_\beta^2 \lambda_6  + \frac{1}{2} v^2 c_\beta^2 \lambda_7 \\
m_{a _{13}}^2 &= c_\beta \rho_5 \\ 
m_{a _{22}}^2 &=   t_\beta \rho_1  - v^2 s_\beta^2 \lambda_5 - \frac{1}{2} v^2 s_\beta^2 t_\beta \lambda_6  - \frac{1}{2} v^2 s_\beta c_\beta \lambda_7  \\ 
m_{a _{23}}^2 &= - s_\beta \rho_5  \\
m_{a _{33}}^2 &= m_{a_{33}}^2
\end{align}
where for convenience we've defined the $\beta$ independent parameters:
\begin{align}
\rho_1 &= \mu_{12}^2 - v_S \kappa_3 - v_S \kappa_4 - v_S^2 \lambda_{10} - v_S^2 \lambda_{11} - v_S^2 \lambda_{17}\\
\rho_2 &= v \kappa_3 + v \kappa_4 + 2 v v_S \lambda_{10} + 2 v v_S \lambda_{11} + 2 v v_S \lambda_{17} \\
\rho_3 & = 2 v \kappa_1 + 4 v v_S \lambda_8 + 2 v v_S \lambda_{14} \\
\rho_4 & = 2 v \kappa_2 + 4 v v_S \lambda_9 + 2 v v_S \lambda_{15} \\
\rho_5 &=  v \kappa_3 - v \kappa_4 + 2 v v_S \lambda_{10} - 2 v v_S \lambda_{11},
\end{align}
where $\tan \beta \equiv v_1 / v_2$.

The mass matrices can be diagonalized by a unitary transformation. We can relate the CP even mass eigenstates to the gauge eigenstates via the coefficient matrix:

\begin{align}
\left( \begin{array}{ccc}  N_{11} & N_{12} & N_{13} \\ N_{21} & N_{22} & N_{23} \\ N_{31} & N_{32} & N_{33}  \end{array} \right)\left( \begin{array}{c} \phi_1 \\ \phi_2 \\ \phi_3 \end{array}\right)=\left( \begin{array}{c} \phi_1^0 \\ \phi_2^0 \\ \varphi_S \end{array}\right),
\end{align}
and also define the the inverse coefficient matrix $N^\prime \equiv N^{-1}$ with components $N^\prime_{ij}$. We'll take $\varphi_1 \equiv h$ to be the Standard Model 125 GeV Higgs. 

Similarly for the CP odd case we define:

\begin{align}
\left( \begin{array}{ccc}  A_{11} &A_{12} & A_{13} \\ A_{21} & A_{22} & A_{23} \\ A_{31} & A_{32} & A_{33}  \end{array} \right)\left( \begin{array}{c} a_1 \\ a_2 \\ Z^{||}   \end{array}\right)=\left( \begin{array}{c} A_1 \\ A_2 \\ A_S \end{array}\right),
\end{align}
One of the states will be eaten by the $Z$ and there will be two physical pseudoscalars remaining.

\end{document}